\documentclass[useAMS,usenatbib]{mn2e}
\usepackage{graphicx,psfig,lscape}

\title[Magnetic Fields in Massive Star Forming Regions]{Magnetic
Fields in Massive Star Forming Regions} \author[R. L. Curran et
al.]{R. L. Curran$^{1,2}$\thanks{E-mail: rcurran@cp.dias.ie}, and
A. Chrysostomou$^{2,3}$ \\ $^{1}$School of Cosmic Physics, Dublin
Institute for Advanced Studies, 5 Merrion Square, Dublin 2, Ireland\\
$^{2}$School of Physics, Astronomy \& Maths, University of
Hertfordshire, College Lane, Hatfield, Herts., AL10 9AB, UK\\
$^{3}$Joint Astronomy Centre, 660 N. A'ohoku Place, University Park,
Hilo, Hawaii 96720, U.S.A.}

\begin{document}

\date{Accepted ------; Received ------}

\pagerange{\pageref{firstpage}--\pageref{lastpage}} \pubyear{2005}

\maketitle

\label{firstpage}

\begin{abstract}
We present the largest sample of high-mass star-forming regions
observed using submillimetre imaging polarimetry. The data were taken
using SCUBA in conjunction with the polarimeter on the JCMT in
Hawaii. In total, 16 star forming regions were observed, although some
of these contain multiple cores. The polarimetry implies a variety of
magnetic field morphologies, with some very ordered fields. We see a
decrease in polarisation percentage for 7 of the cores. The magnetic
field strengths estimated for 14 of the cores, using the corrected CF
method, range from $<$0.1 mG to almost 6 mG. These magnetic fields are
weaker on these large scales when compared to previous Zeeman
measurements from maser emission, implying the role of the magnetic
field in star formation increases in importance on smaller
scales. Analysis of the alignment of the mean field direction and the
outflow directions reveal no relation for the whole sample, although
direct comparison of the polarimetry maps suggests good alignment (to
at least one outflow direction per source) in 7 out of the 15 sources
with outflows.
\end{abstract}


\section{Introduction}

One of the remaining problems yet to be understood within star
formation is the collapse of a cloud into a star, and the relatively
slow star formation rate observed. This problem suggests there is some
form of support preventing the clouds from collapse (at least
initially). For instance, if typical (M $\sim$ 10$^{3-4}$ M$_{\odot}$)
molecular clouds had no support, and so were in free-fall collapse,
with all the mass going into stars, cloud lifetimes would be
unrealistically short, and the star formation rate would be up to
three orders of magnitude greater than observed
\citep{mous76}. Thermal pressure is weak compared to the gravitational
stresses in the cloud, so the support may come from the magnetic field
that permeates the gas or the pressure of turbulent eddies -- indeed
it is likely that these two mechanisms are coupled.

Spinning elongated dust grains in such star forming regions can become
aligned to the local magnetic field, such that the grains' semi-major
axes are perpendicular (on average) to the magnetic field lines. The
net thermal emission from such grains is polarised (perpendicular to
the direction of the magnetic field lines), and can be used to trace
the morphology of the magnetic field as seen projected onto the plane
of the sky. Further information can be gained on the magnetic field
via Zeeman splitting \citep{sarma}, which yields the line of sight
magnetic field strength. Ion/neutral linewidths \citep{houde1}, can
also be used to gain information on the orientation of the magnetic
field.

In order to understand the importance of the magnetic field on large
scales, consistent with tracing the magnetic field in the envelope of
these cores, we present SCUBA 850$\mu$m imaging polarimetry of 20 star
forming cores. We also introduce a new, improved (more representative)
method of data reduction for SCUBA polarimetry, which presents the
errors in polarisation in a truer light. We discuss the implied
magnetic field morphology across the cores, as well as the estimated
field strength for some of the regions using the modified \citet[][
henceforth CF]{cf} method. We use this estimate to indicate the
importance of the magnetic field on these scales on the support of
these clouds, and compare the mean magnetic field direction to the
outflow directions, to see if they correspond in the plane of the sky.

\section{Observations \& Data Reduction}

The sample of high-mass star-forming regions were selected from the
Arcetri water maser atlas \citep{comoretto,brand} based on the
following criteria:
i) well known and studied star-forming regions, ii) varied
morphologies, iii) submillimetre bright \citep{timscott};
and iv) observable from the JCMT in Hawaii (i.e. $-40^{\circ}
\leq \delta \leq +70^{\circ}$).

The observations took place over several nights between 1998 May 16
and 2000 October 10 at the James Clerk Maxwell Telescope (JCMT) in
Hawaii. The Submillimetre Common User Bolometer Array (SCUBA)
\citep{holland} was used in conjunction with the polarimeter --
consisting of a rotating quartz half-wave retarder ahead of a fixed
wire-grid analyser; \citet{jane} -- which was mounted on the entrance
window to SCUBA. The Jiggle-mapping mode of observation was used,
involving `jiggling' the secondary mirror to fully sample the
beam. Sixteen different positions (1 second exposures at each) are
required to fully sample the map -- these positions have a separation
of 6.18\arcsec\, for the Long-Wavelength (850$\mu$m) array. The
secondary mirror performed the usual chop during the jiggle pattern to
provide atmospheric cancellation. The nod was carried out over periods
of around 10-20 seconds to eliminate slowly varying sky gradients.

The polarimeter complicates the normal observing procedure such that
complete 16-point jiggle maps are required at specific positions of
the half-wave retarder, each separated by 22.5\degr. Therefore 16
jiggle maps are observed to complete one cycle of the retarder. 
The direction of the chop-throw was decided based
on the morphology of the target. 

The data reduction was carried out using the routines from the SCUBA
User Reduction Facility (SURF) \citep{jennesslight} to reduce the
SCUBA images and routines from POLPACK \citep{berry} were used to
reduce the polarimetry. The nod and chop of the telescope were
corrected for and the flat-field applied to the observations in the
standard manner. The atmospheric extinction was calculated based on
the start and end times of each observation: throughout the night the
Caltech Submillimetre Observatory (CSO) phase monitor measures the
$\tau_{225 GHz}$ (or $\tau_{CSO}$), and a polynomial was fitted to the
measured points. This polynomial was then applied to the start and end
times of the observation to calculate the $\tau_{CSO}$ for the
observation and converted into $\tau_{850 \mu m}$ using:

\begin{equation}
\tau_{850 \mu m} = 3.99 \times (\tau_{CSO} - 0.004)
\end{equation}

\noindent for the data taken before October 2000, and:

\begin{equation}
\tau_{850 \mu m} = 4.02 \times (\tau_{CSO} - 0.001)
\end{equation}

\noindent for the data taken during and after October 2000. The
extinction is assumed to vary linearly throughout the observation. The
airmass at which each bolometer measurement was made is calculated
then multiplied by the zenith sky extinction. Each data point was then
multiplied by the exponential of the optical depth to give the value
that would have been measured in the absence of the
atmosphere. Bolometers which were deemed excessively noisy were
switched off at this point in the data reduction.  The sky noise was
removed using bolometers that had no significant flux from the
source. The average (mean) flux from these bolometers was assumed to
come from sky emission, and was subtracted from all of the bolometers
in order to remove the sky signal. The instrumental polarisation was
removed in the usual manner. The data were re-gridded using a Gaussian
weighting function, with the scale set to 7\arcsec\, (half the
beamsize). The pixel size was set to 6.18\arcsec\,, matching the
jiggle pattern, in order to calculate the polarimetry vectors
accurately.

POLPACK packages are then used to calculate the Stokes parameters by
fitting the following curve \citep{axon} to the data:

\begin{equation}
I'_{k} = \frac{t}{2}(I + \epsilon(Q\cos2\phi_{k} + U\sin2\phi_{k}))
\end{equation}

\noindent where $I'_{k}$ is the expected intensity in image $k$, $t$
is the wire-grid analyser transmission factor, $\epsilon$ is the
analyser polarising efficiency factor and $\phi_{k}$ is the effective
retarder position angle after correction for the parallactic angle for
image $k$. The polarisation percentage and position angles of the
vectors are then calculated from the Stokes parameters, without any
binning. The catalogue was clipped such that noisy polarisation
vectors were not included. The clipping used was $I>0$ and
$dP<0.75$. For 3\% polarisation, this represents a maximum position
angle error of 7.15\degr. It was chosen to clip the vectors on
polarisation errors instead of signal-to-noise, as clipping on the
latter would result in disposing of points where the polarisation is
low or zero, both of which are perfectly valid measurements.

Finally, flux calibration \citep{tim02} was carried out using
observations of Uranus, Saturn and CRL2688. The flux calibration
factors (FCFs) are listed in table \ref{tab:fcf}.  These FCFs were
applied to the data, with any objects observed over more than one
night having the data for each night reduced separately, then the flux
calibrated images co-added to create the final image.

\begin{table}
\centering
\caption{Table of flux calibrators and flux calibration factors.} 
\label{tab:fcf}
\begin{tabular}{l|c|c}
\hline
Date & Flux Calibrator & FCF \\
  &  & Jy/arcsec$^{2}$\,/V \\
   \hline
19980516 & Uranus & 2.35 \\
19990620 & Uranus & 2.02 \\
19990705 & Uranus & 2.21 \\
19990706 & Saturn & 1.81 \\
20001006 & Uranus & 1.96 \\
         & CRL2688 & 1.72 \\
20001010 & CRL2688 & 1.71 \\
  \hline
\end{tabular}
\end{table}

\subsection{Pixel Size}
\label{pix}

Since the introduction of SCUBA on the JCMT, it has become
``standard'' to reduce 850$\mu$m data with a pixel spacing of
3.09\arcsec\,. This is mainly due to a 3.09'' jiggle-step being
required to fully sample the beam at 450$\mu$m. This allowed for
easier comparisons between 450$\mu$m and 850$\mu$m data. The method of
data reduction naturally continued when the polarimeter was
introduced.

Careful inspection of data reduced by this method (pixel size set to
3.09\arcsec and with 2$\times$2 binning of vectors)
and reduced using a pixel size of 6.18\arcsec with no binning, revealed
significant differences in the number of vectors. The polarisation
errors in the latter reduction were higher than those in the first
reduction method, resulting in a decrease in the number of vectors
selected after error clipping.

The change in the errors in polarisation are due to the binning. When
the pixel size is set to 3.09\arcsec, four pixels are produced for
every one when the size is 6.18\arcsec. A vector is then produced for
each pixel, with the vectors for the smaller pixels being binned
together --- adding the polarisations together in quadrature to
calculate the error. This erroneously reduces the resultant errors.

The observing mode used means that one independent polarisation measurement is
taken at steps of every 6.18\arcsec\,, the results of which can be
superimposed over an intensity image created with 3.09\arcsec\, pixels.

\begin{table*}
\centering
\caption{The sources with their observed parameters. Magnetic field strengths are not calculated for all cores, e.g. DR21(OH) Main. The vectors across the cores which have no field strength estimates have low polarisation signal-to-noise (see section 3.2).} 
\label{tab:param}
\begin{tabular}{l|c|c|c|c|c|c|c|c|c}
\hline
Core & RA & DEC & Distance & Temp & Flux (850$\mu$m) & Mass & $n$ & B & Temp. Ref.$^{a}$ \\
  & (hh:mm:ss) & (\degr \arcmin\ \arcsec\ ) & (kpc) & K & (Jy) & (M$_{\sun}$) & cm$^ 
{-3}$ & mG & \\
   \hline

Cepheus A       & 22 56 17.9 &  62 01 49 & 0.725 & 35 &117 &  200 &  
7.9 $\times$ 10$^{5}$ & 5.7 & 1 \\
DR21(OH)N       & 20 38 59.5 &  42 23 31 & 3     & 20 &41$^{b}$ &  
2500 & 1.6 $\times$ 10$^{5}$ & 1.1 & 2 \\
DR21(OH) Main   & 20 39 00.8 &  42 44 49 & 3     & 58/30 &60$^{c}$ &  
900/2100 & 1.8/4 $\times$ 10$^{5}$ & & 2 \\
GGD27           & 18 19 12.1 & -20 47 31 & 1.7   & 20 &34 &  660 &  
2.1 $\times$ 10$^{5}$ &  0.2 & \\
GL2136          & 18 22 29.1 & -13 29 46 & 2     & 28 &28 &  470 &  
9.1 $\times$ 10$^{4}$ & 0.3 & 3 \\
GL2591          & 20 29 24.9 &  40 11 21 & 1.5   & 28 &34 &  320 &  
1.5 $\times$ 10$^{5}$ & 0.5 & 3 \\
GL437           & 03 07 23.7 &  58 30 50 & 2     & 20 &15 &  400 &  
7.8 $\times$ 10$^{4}$ & & \\
IRAS 20126+4104 & 20 14 26   &  41 14 42 & 1.7   & 27 &26 &  330 &  
1.1 $\times$ 10$^{5}$ & 0.1 & 3 \\
IRAS 20188+3928 & 20 20 39.3 &  39 37 52 & 4     & 39 &48 & 2100 &  
5.1 $\times$ 10$^{4}$ &  0.2 & 4 \\
L1287           & 00 36 47.5 &  63 29 02 & 0.85  & 34 &24 &   60 &  
1.4 $\times$ 10$^{5}$ & & 5 \\
MonR2           & 06 07 46.3 & -06 23 09 & 0.95  & 50 &52 &  100 &  
1.7 $\times$ 10$^{5}$ &  0.2 & 6 \\
NGC6334A        & 17 20 18.6 & -35 54 45 & 1.7   & 50 &149 & 870 &  
2.8 $\times$ 10$^{5}$ &   0.9 & 7 \\
NGC6334AE       & 17 20 23.9 & -35 54 55 & 1.7   & 50 & 75 & 440 &  
1.4 $\times$ 10$^{5}$ &  & 7 \\
RCrA            & 19 01 53.6 & -36 57 07 & 0.129 & 20 & 44 &   5 &  
3.6 $\times$ 10$^{6}$ &  1 & \\
S140            & 22 19 18.1 &  63 18 49 & 0.9   & 27 &72 &  260 &  
5.5 $\times$ 10$^{5}$ & 0.4 & 3 \\
S146            & 22 49 29.1 &  59 54 53 & 4.7   & 42 &17 & 1150 &  
1.3 $\times$ 10$^{4}$ & 0.1  & 8 \\
S146N           & 22 49 30.9 &  59 55 30 & 4.7   & 20 &5$^{c}$&  900$^ 
{c}$ & 3.4 $\times$ 10$^{4}$ & & \\
S157            & 23 16 04   &  60 02 06 & 2.5   & 20 &21 &  880 &  
8.8 $\times$ 10$^{4}$ &  0.2  &  \\
W49NW           & 19 10 13.2 &  09 06 14 & 11.4  & 45 &288 &86000&  
9.1 $\times$ 10$^{4}$ &  $<$0.1 & 9 \\
W49SE           & 19 10 21.8 &  09 05 03 & 11.4  & 45  &61 &18200&  
1.9 $\times$ 10$^{4}$ & & 9 \\
  \hline
\end{tabular}
\flushleft $^{a}$ Where no reference is given, temperatures are assumed to be 20 K, consistent with the coldest measured temperature in this sample, DR21(OH). This therefore leads to upper limits for the masses of these objects.\\
$^{b}$ For a cylinder of radius 0.33pc (22.7\arcsec\,) and
height 0.93 pc (64\arcsec\,).\\ 
$^{c}$ For a 40\arcsec aperture, instead of a 60\arcsec\,, to stop contamination between sources.\\
Temperature references are: (1) \citet{bottinelli}; (2) \citet{mangum92}; (3) \citet{vandertak}; (4) \citet{mccutcheon}; (5) \citet{sandellweintraub}; (6) \citet{thronson}; (7) \citet{sandell}; (8) \citet{wu05}; (9) \citet{harvey77}
\end{table*}
\normalsize

\section{Core Properties}
\subsection{Core Masses}

The mass of the star-forming cores can be calculated by assuming:

\begin{equation}
M_{\rm Total} = \frac{gS_{\rm \nu}d^{2}}{\kappa_{\rm \nu}B_{\rm
\nu}(T_{\rm dust})}\label{eq:mass}
\end{equation}

\noindent where $g$ is the gas-to-dust ratio, $S_{\rm \nu}$ is the flux,
$d$ the distance, $\kappa_{\rm \nu}$ the absorption coefficient at
frequency $\rm \nu$, and $B_{\rm \nu}(T_{\rm dust})$ is the Planck  
function for
frequency $\rm \nu$ at a temperature of $T_{\rm dust}$.

Using a gas-to-dust ratio of 100:1 \citep{hildebrand}, and an
absorption coefficient of 0.15 m$^{2}$ kg$^{-1}$ estimated from
\citet{ossen} based on a number density of n$_{H}$ = 10$^{5}$
cm$^{-3}$, thick ice mantles and a formation timescale of 10$^{5}$
years, mass estimates for the cores were calculated (listed in
table~\ref{tab:param}, along with the other parameters derived from
the observations).

The mass of these objects are subject to uncertainties within the used
parameters, especially any errors in measuring the distance to the
core. The gas-to-dust ratio, which may be as low as 45:1
\citep{mccutcheon} also introduces an error that may be up to a factor
of 2. The absorption coefficient at 850$\mu$m still has not been
determined precisely \citep{hildebrand,chini}, although the value
adopted in this paper from \citet{ossen} agrees with the values
determined by \citet{bianchi} and \citet{visser}. The masses we
determine in this paper are therefore the upper limits for the
temperature of the cores. 

\subsection{The Magnetic Field Strength}
\label{cf}

The CF relation can be used to obtain the plane of the sky average
magnetic field strength. This method is based on equipartition and the
ability of the magnetic field to retain straight field lines under the
influence of turbulence. The plane of the sky average field strength
can be calculated via:

\begin{equation}
\langle B_{\rm pos} \rangle = f \sqrt{4\pi\rho}\frac{\sigma_{v_{\rm  
los}}}{\sigma_{\theta}}\ \ \ G,\label{bfield}
\end{equation}

\noindent where $\rho$ is the mean density (g cm$^{-3}$),
$\sigma_{v_{\rm los}}$ the line of sight velocity dispersion (cm
s$^{-1}$), $\sigma_{\theta}$ is the dispersion in polarisation
position angles (measured east of north) and is corrected for
measurement errors ($\sigma^{2}_{\theta}= \sigma^{2}_{\rm measured} -
\sigma^{2}_{\rm error}$) where $\theta$ is in radians, and $f$ is a
correction factor found to be $\sim$0.5 \citep{heitsch}.

Errors on the position angle of the vectors are calculated based on
the polarimetry signal-to-noise:

\begin{equation}
d\theta = \frac{28.6^{\circ}}{s_{\rm p}}
\end{equation}

\noindent where $d\theta$ is the error in the position angle and
$s_{\rm p}$ is the signal-to-noise in the polarisation.  Therefore in
regions where the polarisation percentage is low (e.g. across the main
core of DR21(OH)), then $\sigma_{meas} \sim \sigma_{err}$ resulting in
misleadingly small dispersion angles, and very strong magnetic
fields. By only selecting the vectors which have a signal-to-noise (in
polarisation) of $>$ 3, this problem is avoided, therefore all of the
magnetic field strength estimates in this paper are based on vectors
selected in this way. Please note, however, as mentioned earlier, the
vectors shown in the figures are not selected in this way, as in terms
of magnetic field morphology, polarisation nulls and low levels of
polarisation are valid detections.

The CF method should be used with caution when calculating the
magnetic field strength in the plane of the sky. The 15\arcsec\,
beam-size of the JCMT means that the small-scale tangling of the
magnetic field field can occur within the beam, and so the measured
vectors only represent the net magnetic field direction, leading to
over-estimates of the field strength. Modelling studies of this effect
\citep{heitsch} have lead to the introduction of a correction factor
$f$, which has been found to be $\sim$ 0.5 (see
eq.~\ref{bfield}). Also, any underlying magnetic field morphology
(e.g. intrinsic field curvature) has not been accounted for in
calculating the dispersion in position angles of the polarimetry
vectors (i.e. this technique assumes \emph{a priori} that the magnetic
field is uniform).

Uncertainties in our estimations of the magnetic field strength arise
from calculating the density of the cores, which incorporates the
errors involved in calculating the mass, they therefore represent
the upper limits of the magnetic field strength. There are also
errors in calculating the volume of the core which contribute as a
spherical geometry has been assumed for each core (except where
otherwise stated), although with no density tracer information, we
lack the ability to modal the three-dimensional structure of the
cores, therefore the upper limits stated are solely for spherical
geometry. The velocity of the gas within the core introduces another
error as a FWHM of $\sim$ 2 km\,s$^{-1}$ has been used but it may be
anywhere between 1 km\,s$^{-1}$ and 3 km\,s$^{-1}$
\citep{brand,thompson}. Measurement errors are also introduced by the
angle $\theta$, although these are relatively small in comparison to
the other errors stated.

\section{Individual Cores}

\subsection{Cepheus A}

\begin{figure*}
\includegraphics[height=22cm]{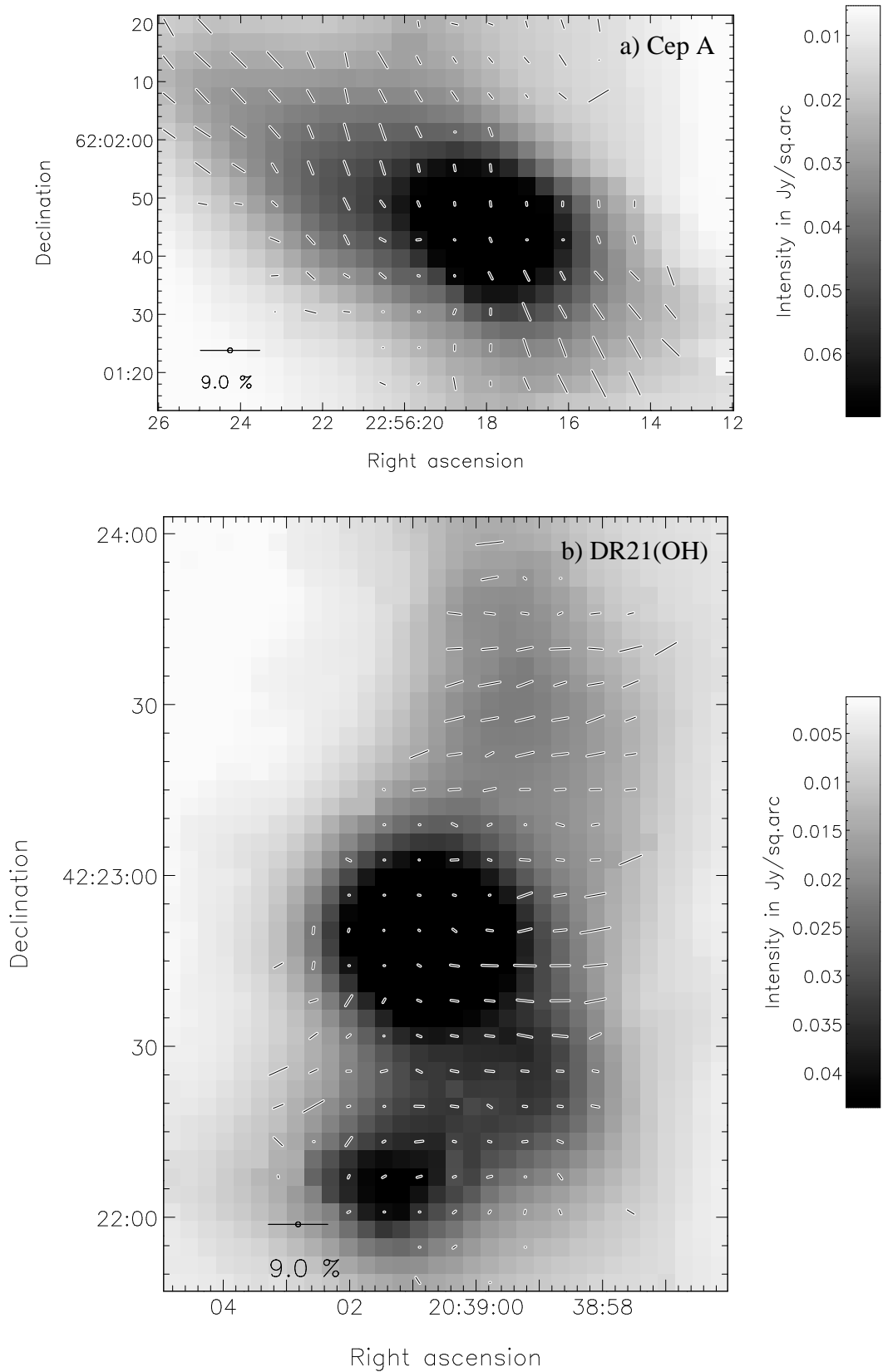}
\caption{Final result of the data reduction method outlined in section
   2. Epoch J2000. The greyscale is the total thermal continuum
   emission at 850$\mu$m, with the {\bf B}-vectors overlaid to
   represent the plane of the sky magnetic field direction. The
   polarisation percentage scale is presented in the lower left of each
   image, with the total intensity scale to the right of each image.}
\label{fig1}
\end{figure*}
\begin{figure*}
\includegraphics[height=22cm]{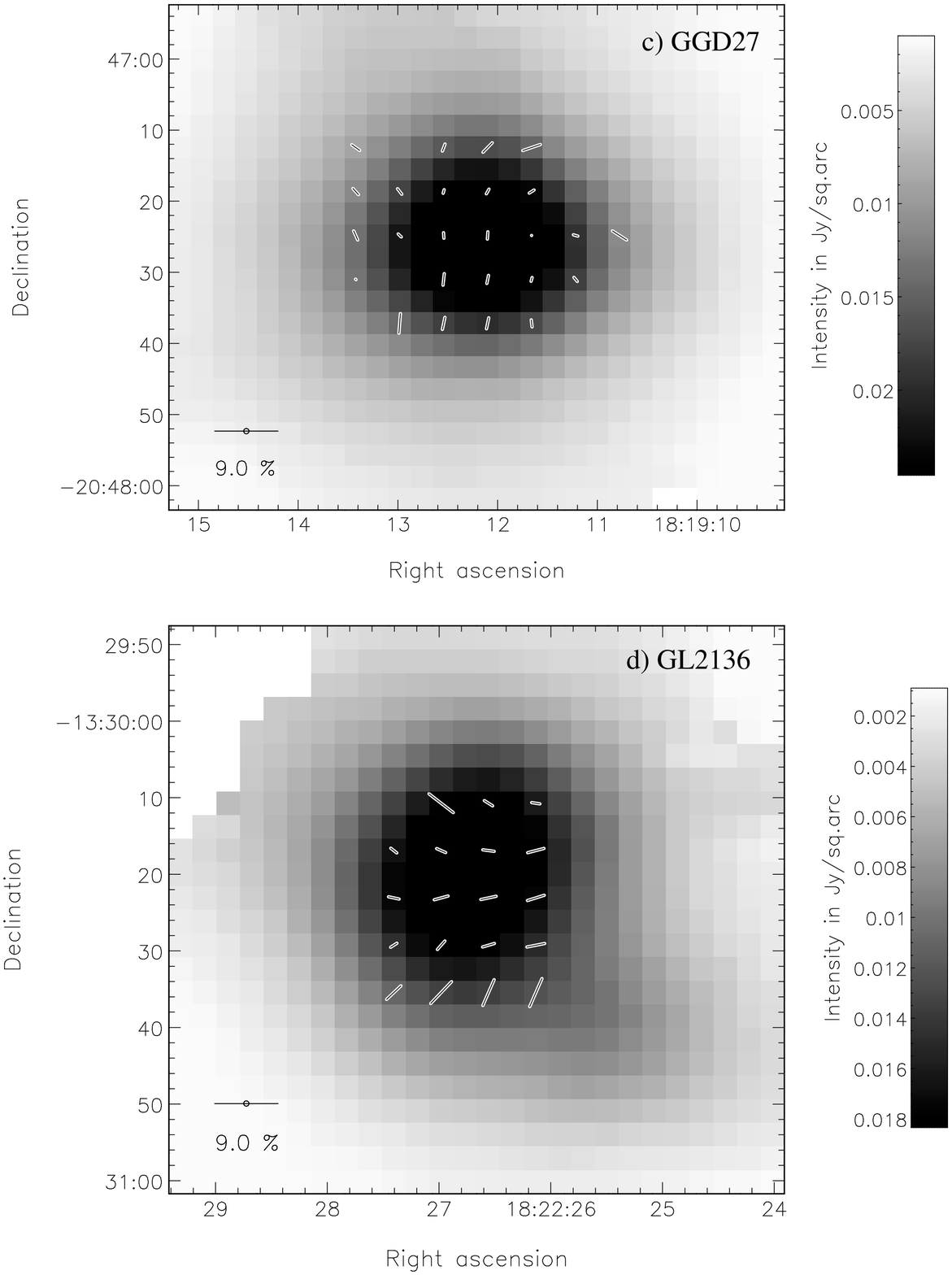}
\begin{center}
Figure~\ref{fig1} (continued)
\end{center}
\end{figure*}
\begin{figure*}
\includegraphics[height=22cm]{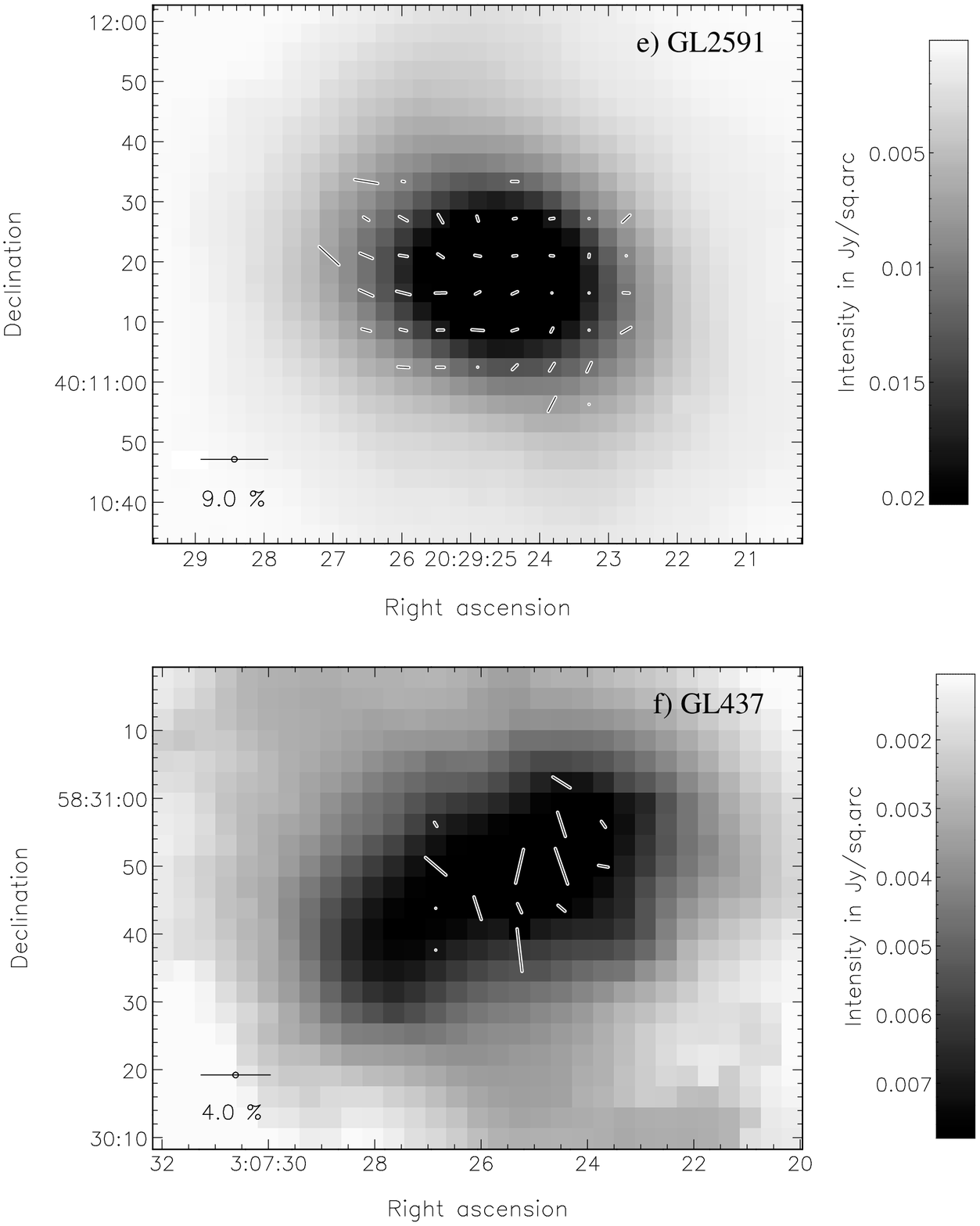}
\begin{center}
Figure~\ref{fig1} (continued)
\end{center}
\end{figure*}
\begin{figure*}
\includegraphics[height=22cm]{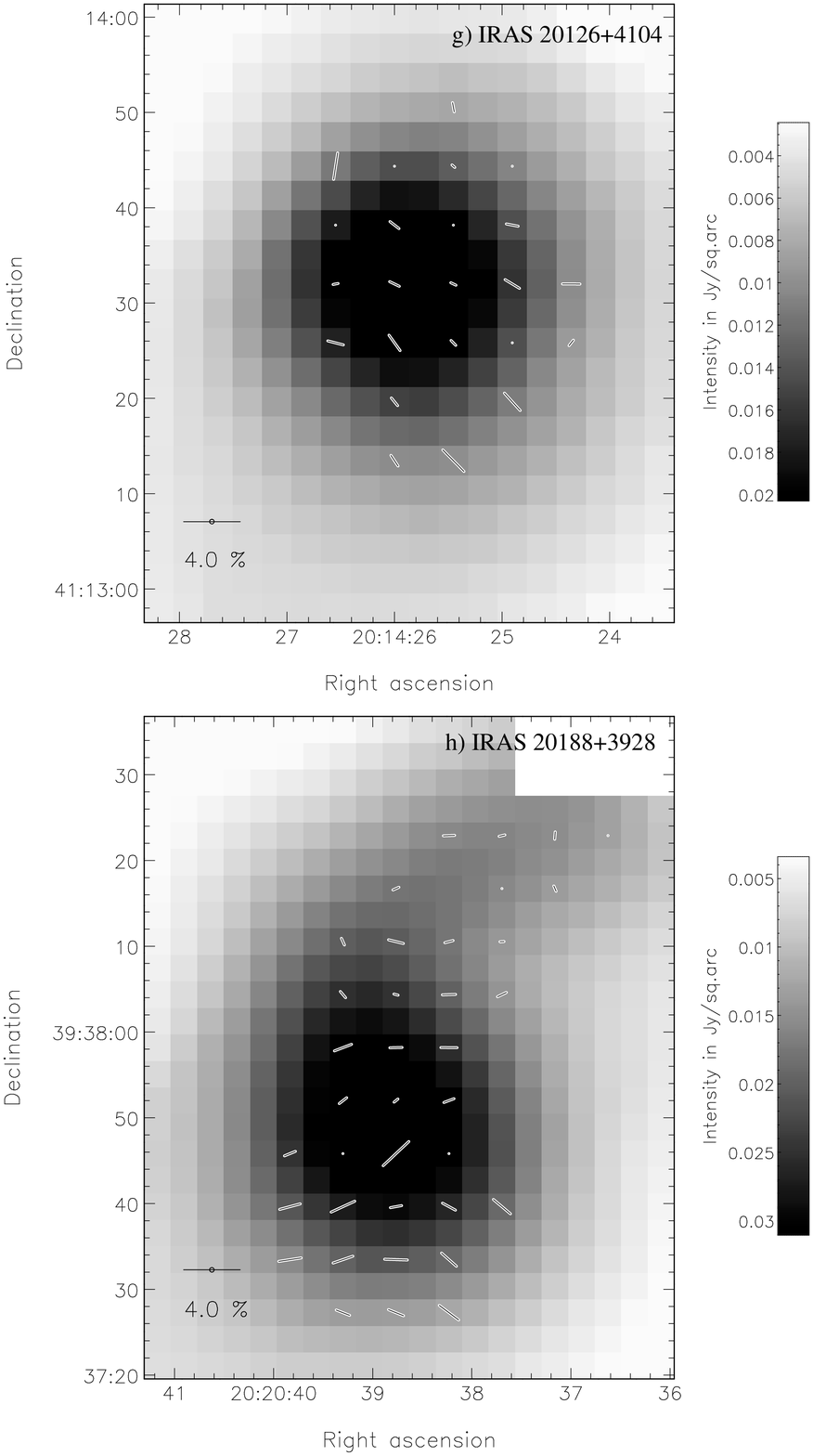}
\begin{center}
Figure~\ref{fig1} (continued)
\end{center}
\end{figure*}
\begin{figure*}
\includegraphics[height=22cm]{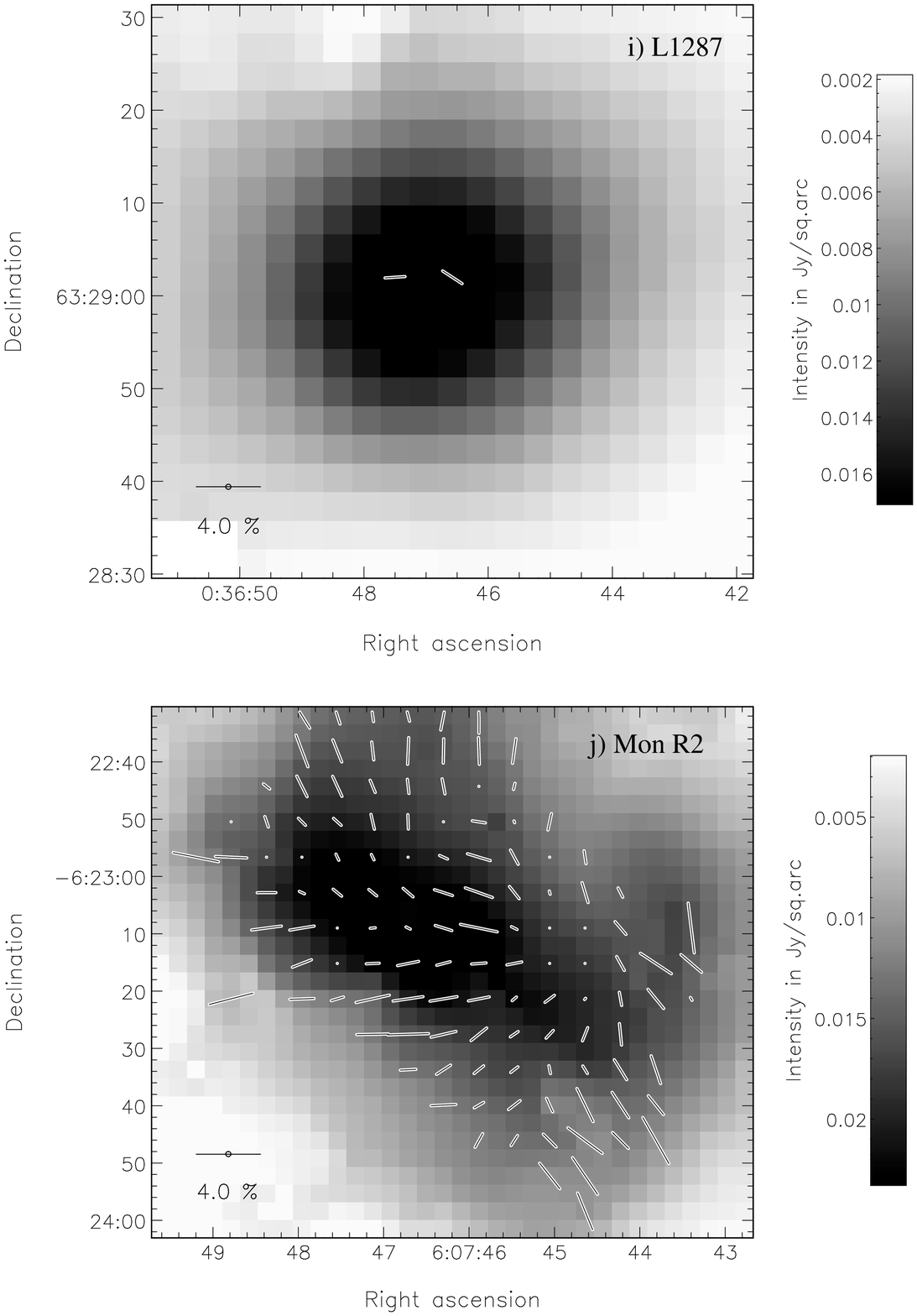}
\begin{center}
Figure~\ref{fig1} (continued)
\end{center}
\end{figure*}
\begin{figure*}
\includegraphics[height=22cm]{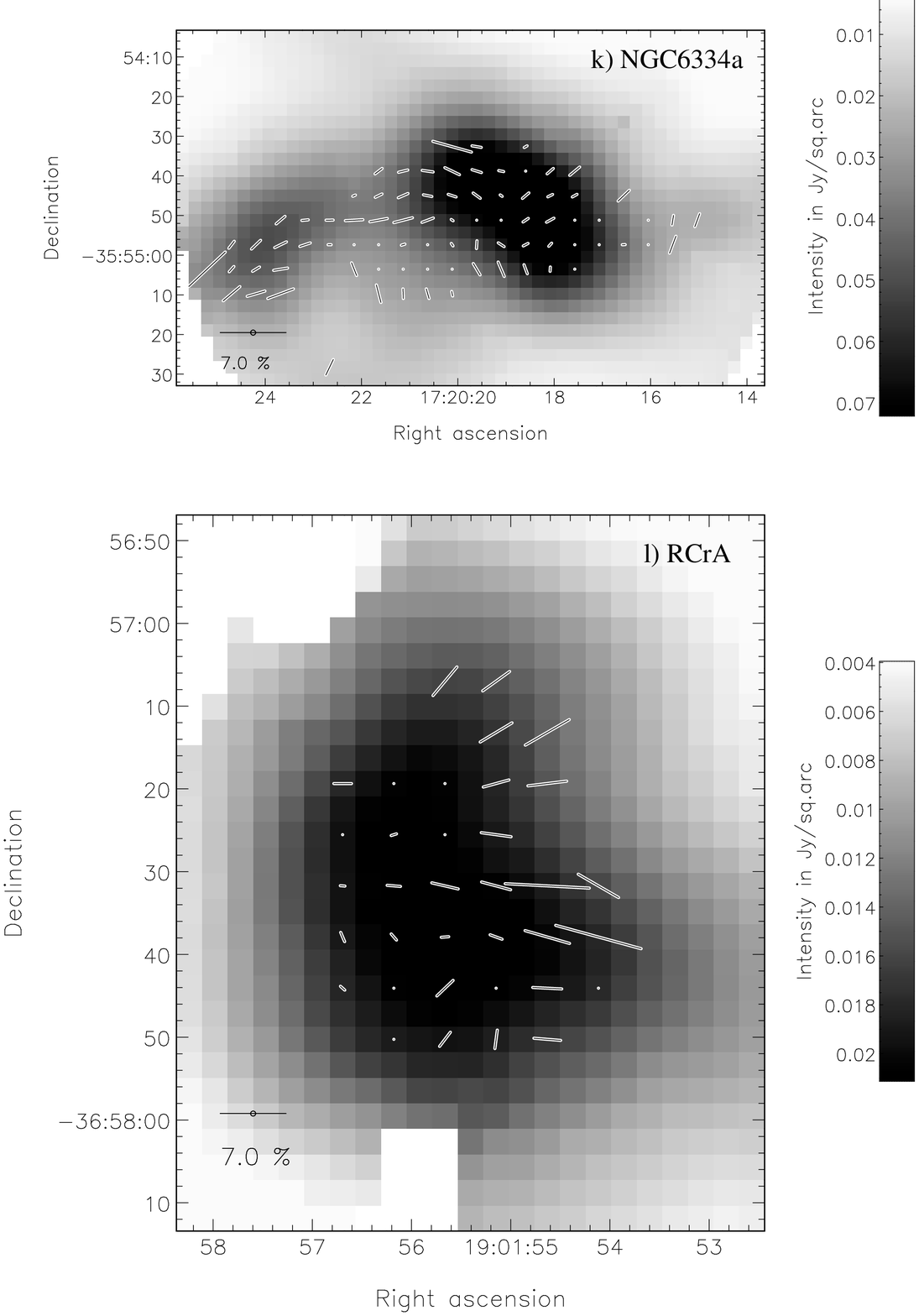}
\begin{center}
Figure~\ref{fig1} (continued)
\end{center}
\end{figure*}
\begin{figure*}
\includegraphics[height=22cm]{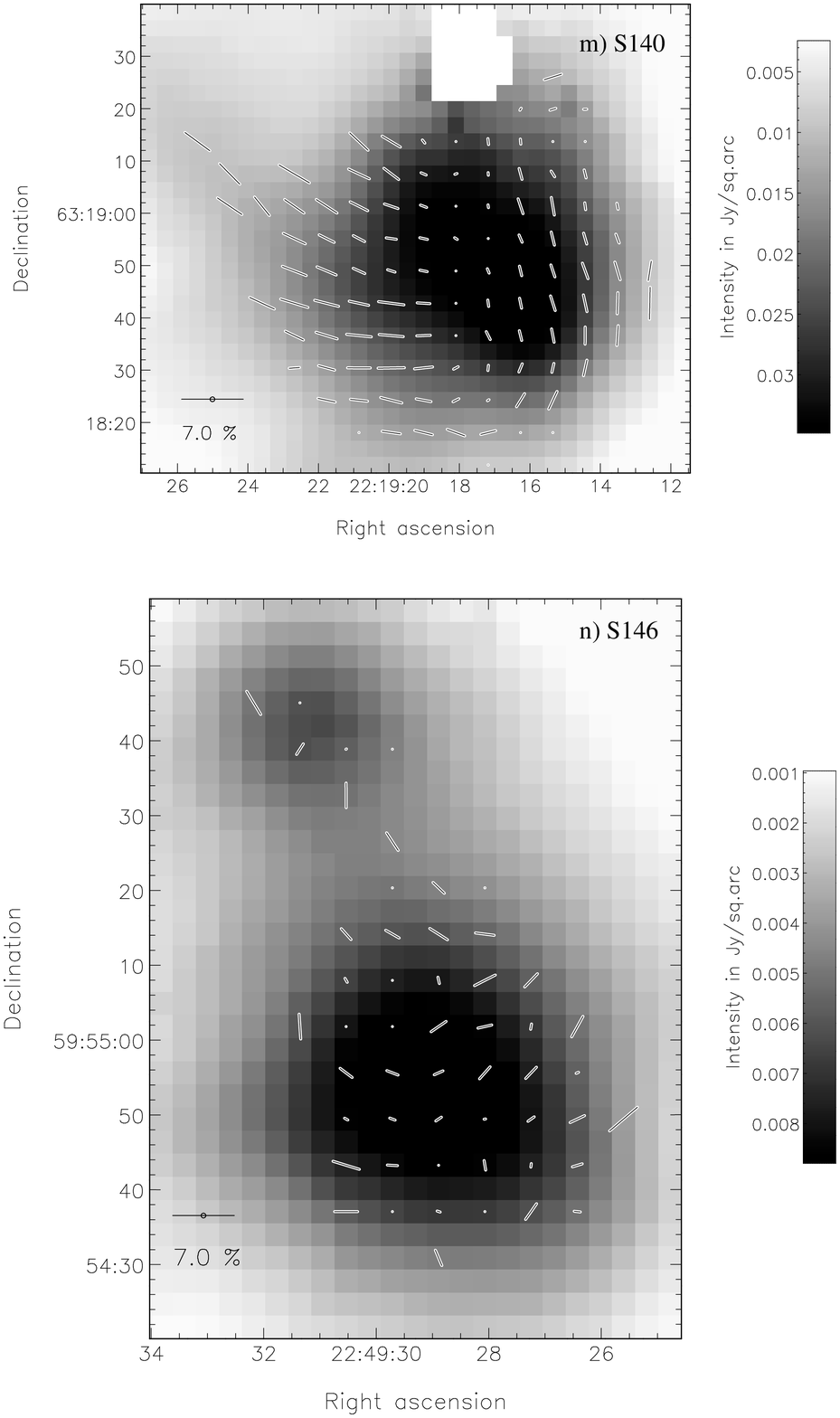}
\begin{center}
Figure~\ref{fig1} (continued)
\end{center}
\end{figure*}
\begin{figure*}
\includegraphics[height=22cm]{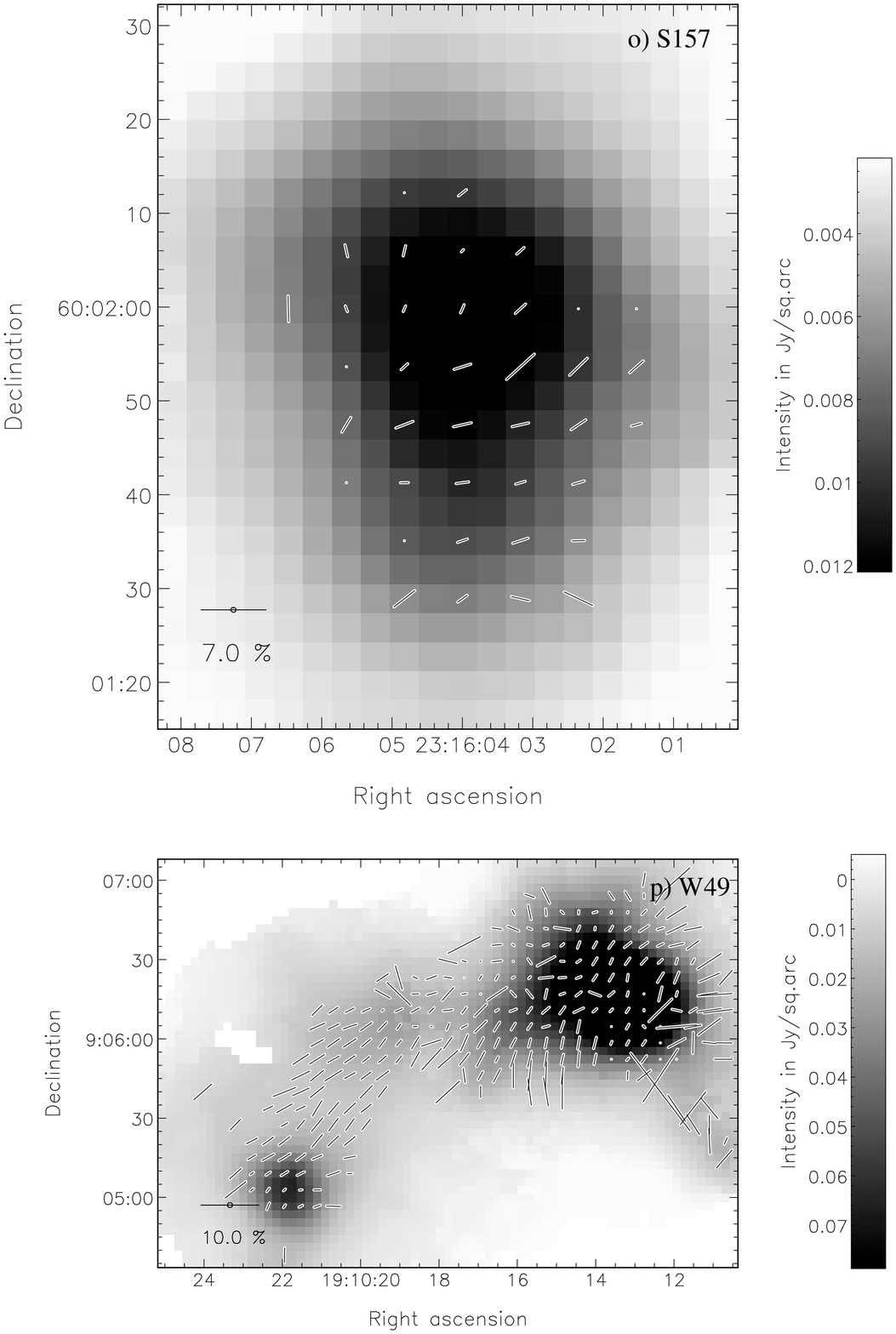}
\begin{center}
Figure~\ref{fig1} (continued)
\end{center}
\end{figure*}

Cepheus A is a well-known star forming region. Located at a distance
of $\sim$730 pc \citep{blaauw}, it is the closest star-forming region
in the sample. The region contains many signatures of massive star
formation such as: a) a sub-mm source (observed here) with a total
luminosity of 2.5 $\times$ 10$^{4}$ L$_{\odot}$
\citep{koppenaal,evans81}, b) a cluster of 14 compact radio continuum
sources \citep{beichman79,hugheswater}, c) clusters of H$_{2}$O and OH
masers \citep{lada81,migenes}, d) extended shock-excited H$_{2}$
emission at 2.12$\mu$m \citep{ballylane,doyon} and e) highly
blueshifted Herbig-Haro objects with large proper motions pointing
away from the activity centre
\citep{lenzenhod,hartiganlada,hartigan86,lenzen}. The well known
outflow in this region has a complex morphology. On scales of $\sim$ 1
pc, the outflow is in the east-west direction, but on smaller scales,
there is an ionised jet that is oriented
northeast-southwest. \citet{narayananwalker96} argue that this may be
due to multiple outflow episodes. They hypothesise that the older
(3--20 $\times 10^{4}$yr) high-velocity outflow is oriented along the
east-west direction, while the extremely high velocity lobes of the
current flow ($\leq 5 \times 10^{3}$ yr) extend to the northeast and
southwest. On even smaller scales, there is evidence for a thermal
radio jet with position angle $\sim$ 48\degr\,, which is perpendicular to a band
of water maser positions that are thought to trace a circumstellar
disc \citep{gomez}.

The magnetic field in Cepheus A has been studied recently using Zeeman
effect of H$_{2}$O masers \citep{Vlemmings06} and OH masers
\citep{Bartkiewicz}. \citet{Vlemmings06} find that the maser emission
associated with the HW2 high-mass young stellar object in this region,
which is argued to trace the circumstellar disc, has a field strength
of 100--600 mG. They find other masers further from HW2 to have
magnetic field strengths of 30--100 mG. In all cases the magnetic
field pressure is calculated to be similar to the dynamic pressure,
indicating the magnetic field is strong enough to control the outflow
dynamics of HW2. \citet{Bartkiewicz} find that the magnetic field is
extremely ordered on arcsecond scales, pointed away from us in the
east, and toward us in the west. Their linear polarisation
measurements from these masers reveal that the direction of the
magnetic field in the plane of the sky is $\sim$ 53\degr, or roughly
parallel to the radio jet axis.

The SCUBA polarimetry of Cepheus A (Fig~\ref{fig1}a) reveal magnetic
field vectors that are ordered around a position angle of 50\degr\, in
the northeast of the observed region, and ordered $\sim$ 40\degr\, in
the southwest of the region. There is a band of depolarisation across
the core which has a position angle of $\sim$ 135\degr\,
perpendicular to the inferred magnetic field direction. This plane of
the sky magnetic field morphology is roughly parallel to the
outflow/radio jet axis that has been observed on similar/smaller
scales. This pattern of magnetic field vectors may be interpreted as a
`pinched-in' or hourglass morphology, with the field twisted towards
the centre of the source, where the depolarisation occurs.

CF calculations for the plane of the sky magnetic field strength lead
to estimates of $\sim$ 6 mG. This is the strongest magnetic field
calculated for the cores in this sample, but is smaller than the field
strengths calculated from the maser emission \citep{Vlemmings06}. This
suggests that the magnetic field is weaker on larger scales, becoming
stronger on smaller scales, closer in to the protostar. The data
presented here, in conjunction with the findings of
\citet{Bartkiewicz} and \citet{Vlemmings06}, lead to a magnetic field
that, in the plane of the sky, is northeast-southwest in direction,
with the magnetic field in the northeast pointing away from us, and in
the southwest pointing toward us.

\subsection{DR21(OH)}

DR21(OH) is part of the star forming complex W75, located in the
Cygnus X region, at a distance of $\sim$ 3 kpc \citep{campbell}. Its
relative close proximity has aided in the fact that it is one of the
most studied star forming regions in the Galaxy. The DR21 H\,{\sc ii}
region is $\sim$ 3\arcmin\, to the south of DR21(OH) and contains a
cluster of late type O stars, and one of the most intense outflows
known. DR21(OH) -- also known as W75S or W75S(OH) -- unlike DR21,
consists of young stars in the process of forming, and as such offers
a glimpse of an earlier stage of evolution of star formation. Previous
continuum studies have shown that DR21(OH) is made up of three compact
continuum sources, DR21(OH) Main, DR21(OH)S and DR21(OH)W, all of
which seem to be actively forming stars \citep{mangum91}. Higher
resolution studies confirmed that DR21(OH)Main is in fact composed of
two smaller cores, MM1 and MM2 \citep{woody89}.

The magnetic field of DR21(OH) is reasonably well studied, and there
are Zeeman estimates of the line of sight magnetic field strength for
MM1 and MM2 \citep{crutcher99}. \citet{lai04} used BIMA to gain
information on the magnetic field via polarimetry of both dust and CO,
in order to map the magnetic field morphology in the
plane of the sky. Comparisons of ion/neutral linewidths have also been
used to establish a three-dimensional impression of the magnetic field
\citep{lai03}. The morphology of the field at the resolution sampled
by BIMA indicates that the magnetic field is ordered, thus implying a
strong field. The magnetic field is estimated to have a strength of
$\sim$ 0.4 mG with an inclination of 36$^{\circ}$ to the line of
sight, and a position angle of 105$^{\circ}$ in the plane of the sky
\citep{lai03}. The BIMA data only measure the polarised dust emission
in patches, and do not reveal the magnetic field morphology throughout
the whole region.

The data presented here (Fig~\ref{fig1}b) include our observed data as well
as some archival data for DR21(OH), which were observed on 2002 October
2 \citep{vallee}, which effectively doubles the time on source for
this target. The archival data were subjected to our method of data
reduction as detailed earlier. The data reveal that the DR21(OH)
region is composed (at this resolution) of one main core DR21(OH)
Main. There are also two fainter cores close to the main core, one to
the south-west, DR21(OH)W and one due south,
DR21(OH)S. \citet{mangum92} identified another core, DR21(OH)N, in
NH$_{3}$ emission, although in the SCUBA data presented here it
appears to have the morphology of a ridge extending northwards from
the main core. There are also molecular outflows associated with the
main core emanating in an east-west direction \citep{lai04}.

Our calculated mass of the Main core is somewhat larger than
those of \citet{mangum91}, but they calculated the individual masses
of MM1 and MM2 using OVRO interferometer measurements of the dust
emission. It is to be expected that our calculations reveal higher
masses, given that they are based on single dish measurements and so
include more diffuse dust on larger scales. Comparisons of our
derived total masses with those of \citet{vallee} reveal our masses
are much higher than their estimates, however they have assumed a
temperature of 100 K for both the Main and northern source. We also
use larger apertures for our calculations. Once these two factors have
been taken into account, the mass estimates are consistent.

The polarimetry indicates that the magnetic field is ordered across
the ridge, parallel to the outflow axis. The percentage polarisation
also remains stable (at $\sim$ 3\%) across the ridge. Across the main
core the percentage polarisation drops, most apparently to the
northeast of the core, coincident with MM1. This is observed in
numerous other cores \citep[for example, ][]{ant,brenda,chris} and
could be due to the magnetic field twisting within the JCMT beam, the
grains becoming more spherical in regions of high density, or the
grains being less efficiently aligned in regions of high density.

To the south of the main core, the vectors across DR21(OH)S are more
dispersed in position angle, such that to the southwest of DR21(OH)S, the
vectors have a position angle of $\sim$ 135$^{\circ}$, changing to
$\sim$ 90$^{\circ}$ northeast of the source. The vectors across the
south-western core are $\sim$ 90$^{\circ}$, the same as across the
ridge.

The polarimetry data, in general agree with the findings of
\citet{vallee}, however we see a smaller dispersion in position angles
throughout the region, which may be due to a combination of higher
signal-to-noise observations and a more careful data reduction. The
polarimetry agree with the findings of \citet{lai04}, with the
polarisation nulls coincident with MM1, where Lai et al. found very
little polarisation from dust. The overall change in direction of the
magnetic field across DR21(OH)Main also agrees with the BIMA data --
both the dust and the CO polarimetry -- indicating that the magnetic
field stays ordered on both large and small scales. The observed
polarisation nulls across MM1 may indicate that either MM1 has a
twisted magnetic field in comparison to MM2, or that MM1 is more
centrally condensed than MM2. Both of these would be consistent with
MM1 being the more evolved of the two cores.

The polarimetry across the ridge (northern source) of the DR21(OH)
region indicate an ordered field, which in itself implies a strong
field. The field is (in the plane of the sky) perpendicular to the
north-south ridge, yielding the possibility that collapse has occurred
along the field lines. CF calculations reveal the magnetic field
strength is of the order $\sim$ 1 mG in the plane of the sky. This is
comparable to the field strengths previously calculated for the two
sources in DR21(OH)Main by \citet{crutcher99} and \citet{lai04}, which
were $\sim$ 1 mG in the plane of the sky from BIMA observations and
$\sim$ 0.5 mG in the line of sight from Zeeman measurements. This
indicates that the magnetic field is not only morphologically uniform
throughout this cloud, but the strength is also reasonably uniform on
different scales in the cloud. \citet{vallee} estimated the strength
of the magnetic field across the region as a whole (finding
780$\mu$G), across the main source (finding 780$\mu$G again) and
across the ridge to the north of the main core ($\sim$ 200$\mu$G). Our
estimate of the magnetic field strength across the ridge is consistent
with their estimate across the entire region, but much greater than
their estimate across the ridge, although they only use the four
closest vectors to the source peak to calculate their field strengths,
whereas we use all of the vectors which fall within the aperture used
for the mass and density calculations.

\subsection{GGD27}

GGD27 is located within Sagittarius, on the southwestern edge of a
dark lane that runs in a northwest-southeast direction. It's kinematic
distance has been calculated to be 1.7 kpc \citep{rodriguez} via the
velocity of observed CO. It is a well known site of star formation,
and located within the SCUBA submillimetre core (marginally resolved
by the JCMT beam) are several infrared sources \citep{stecklum}. There
is a large-scale CO north-south outflow (blue lobe in the north) from
an embedded source (IRS2) \citep{yamashita87a} as well as a CS disc
elongated in the east-west direction. \citet{marti93} discovered one
of the largest radio jets emanating from this region -- it's source
being coincident with IRAS18162-2048, an extremely luminous star
(2$\times 10^{4}L_{\sun}$). They found the radio jet has a position
angle of 21\degr\,, extends 5.3 pc, and is extremely collimated, with
an opening angle of just 1\degr.

The polarimetry data of GGD27 (Fig~\ref{fig1}c) imply a magnetic
field that has a north-south direction, roughly parallel to the
molecular outflow. To the north of the core, the polarimetry show that
the magnetic field `fans out' -- the magnetic field vectors to the
north-east of the core have a north-east orientation, whilst the
vectors to the north-west of the core have a north-west
orientation. This could be an indication that the field is pinched in
closer to the core, although no depolarisation is observed across the
core, suggesting that the magnetic field is not sufficiently
twisted. The plane of sky magnetic field strength is estimated to be
$\sim$ 0.2 mG.

\subsection{GL2136}

Near-infrared studies of GL2136 \citep{minchin91,kastner92} indicate
that there is a circumstellar disc or torus roughly at a position
angle of $\sim$ 45\degr\,. A kinematic distance of 2 kpc has been
calculated by \citet{menten04}. Both water and OH maser emission has been
observed, with analysis of the left and right circularly polarised
components of the 1665 GHz feature leading to line of sight magnetic
field estimates of 1 mG \citep{menten04}. CO observations have
revealed a massive (50M$_{\sun}$) molecular outflow from the source,
perpendicular to the observed disc \citep{kastner94}.

The submillimetre continuum image of GL2136 (Fig~\ref{fig1}d) reveals
a marginally resolved object with a small extension to the
southwest. The magnetic field vectors are mainly in the east-west
direction, however in the south of the source they are more
northwest-southeast orientated. The average position angle is
102\degr\,, around 30\degr\, away from the outflow direction. There is
no depolarisation towards the centre of the core. CF calculations
reveal plane of sky magnetic field strengths of $\sim$ 0.3 mG. This is
less than half the value for the line of sight component measured from
maser emission, therefore, either the magnetic field is weaker on
large scales, or, if we envisage the magnetic field wrapping around
the outflow axis (which is close to the plane of the sky -- see
table~\ref{tab:pol}), it must have a large toroidal component to
produce the large line of sight component of the field strength --
indeed, this may explain the relatively large ($\sim$30\degr\,)
difference between the outflow axis and the magnetic field direction.

\subsection{GL2591}

GL2591 is a well-studied source, although its distance is still quite
uncertain. Distance estimates range from 1 kpc \citep{mozurkewich} to
greater than 2 kpc \citep{merrill}. The majority of estimates are
between 1 and 2 kpc \citep[e.g.][]{wendker}, therefore a mid-range
distance of 1.5 kpc is adopted here. \citet{hasegawa} studied the
molecular outflow in detail, finding a small-scale, well-collimated
fast outflow in an east-west direction superposed on the previously
known northeast-southwest, large-scale, less-collimated slow outflow.

\citet{hutawarakorn05} studied the magnetic field using Zeeman
splitting of OH masers. The masers formed an elliptical shape
perpendicular to the outflow axis, which was suggested to be a
molecular torus of radius $\sim$ 750 AU, inclined at 55\degr\, to the
line of sight. The magnetic field was found to range from -1.6 to +3.8
mG, reversing direction on opposite sides of the disc, indicating a
toroidal component. The linear polarisation vectors were found to be
both parallel and perpendicular to the outflow direction.

The SCUBA data (Fig.~\ref{fig1}e) show a marginally resolved
object, and the magnetic field vectors have, in general, an east-west
direction across the core, parallel to the small-scale
(90\arcsec$\times$20\arcsec\,) outflow. The western side of the core
has a polarisation null, and to the south of this, the magnetic field
appears to be in a southeast-northwest direction. To the eastern side
of the core, the vectors begin to curl up towards the northeast. The
average position angle of the polarisation vectors is 95\degr\, aligning with
the outflow axis. The polarisation null to the west of the object
suggests a twisted field, agreeing with the findings of
\citet{hutawarakorn05}. The plane of sky field strength is estimated
to be $\sim$ 0.5 mG. When compared with the estimates of
\citet{hutawarakorn05}, which indicate a strong toroidal field in the
disc, the data imply that the magnetic field, once again, becomes
weaker on larger scales.

\subsection{GL437}

GL437 is located at a distance of $\sim$ 2 kpc \citep{arquilla}. It is
composed of a compact cluster of young stars (including B stars), and
a reflection nebulosity which is centered on WK34
\citep{weintraubkastner96}. There is also a broad molecular bipolar
outflow that extends $\sim$ 1 pc and is orientated roughly north-south
\citep{gomez92}, with the south lobe being blueshifted. The SCUBA data
(Fig.~\ref{fig1}f) reveal that this source is an elongated core
orientated in the northwest-southeast direction. The polarimetry is
aligned such that the magnetic field appears to be projected
perpendicular to the ridge of gas and dust forming the elongated
source, roughly parallel to the outflow axis. There is a polarisation
null to the southeast of the core, and the polarisation percentage
changes seemingly randomly across the core.

\subsection{IRAS 20126$+$4104}

IRAS 20126+4104 is located in the Cygnus-X region, at a distance of
1.7 kpc \citep{wilking}. It is classified as a high mass protostellar
object \citep[HMPO;][]{sridharan02} --- although at a later stage of
evolution than those discussed in ~\citet{hmpos} --- it is the high mass
equivalent of a class 0/I protostar. It has a well known molecular
outflow and an ionised jet. The CO outflow is roughly north-south in
orientation (position angle of 171\degr) with redshifted gas in the
south and blueshifted gas in the north. The jet, detected by emission
knots of H$_{2}$ and [S{\sc ii}] has a position angle of 117\degr\,
\citep{shepherd00}. \citet{shepherd00} conclude that the most likely
interpretation for this is for the collimated jet to be precessing
through an angle of $\sim$45\degr. More recently, \citet{lebron06}
studied the kinematics of this region and concluded that whilst a
precessing jet is supported by the data, multiple flows driven by
independent sources cannot be ruled out.

The polarimetry data (Fig.~\ref{fig1}g) reveal that the projected
magnetic field has, in general a northeast-southwest orientation. This
is roughly at an angle of $\sim$ 40\degr\, to the ionised jet, and
almost perpendicular to the CO outflow, which may be evidence of a
helical field. There are several polarisation nulls to the north of
the core, away from the intensity peak. The plane of sky component of
the magnetic field is small -- only 0.1 mG.

\subsection{IRAS 20188$+$3928}

IRAS 20188+3928 is located in the Cygnus region at an uncertain
distance between 0.4--4 kpc \citep{little}. All further calculations
in this paper assume a maximum distance of 4 kpc, and so all values
quoted are upper limits. \citet{little} observed this region in CO and
HCO$^{+}$, and found a CO bipolar outflow, which has a north-southwest
direction. The redshifted gas is to the north, and the blueshifted gas
is in the southwest. The SCUBA data (Fig.~\ref{fig1}h) shows a
curved ridge of gas and dust extending from the north of the core,
curling around to the west. The polarimetry (in general) indicate that
the magnetic field lines are east-west in orientation, projected onto
the plane of the sky. Towards the southwest of the core, the
polarimetry vectors begin to deviate from the east-west orientation
towards a northeast-southwest orientation, aligning with the outflow
in the southwest.  The degree of polarisation is measured to be (on
average) lower across the ridge than across the brighter core. A plane
of sky field strength component of $\sim$0.2 mG has been calculated.

\subsection{L1287}

L1287 is a dark cloud at a distance of 850 pc \citep{yang}. There is
an energetic outflow associated with the cloud, orientated in a
northeast-southwest direction. At the centre of the outflow there is a
very cold IRAS source IRAS 00338$+$6312 \citep{yang}. The
submillimetre data (Fig.~\ref{fig1}i) reveal a marginally resolved
object. There is little polarimetry (2 vectors) as the integration
time for this object was only 0.85 hrs, and so the errors in
polarisation are large.

\subsection{MonR2}

Monoceros R2 is a site of ongoing star formation, with a compact H{\sc ii}
region \citep{wood89} and H$_{2}$O and OH masers
\citep{downes,knapp}. The outflow in this region is one of the largest
known, with a total extent of 6.8 pc \citep[assuming a distance of 950
pc;][]{racine}. This outflow extends to the north, east and south-west
of the region. \citet{giannakopoulou} assume that the large scale
north-southwest outflow is old, and that the source of this outflow is
now inactive, as the lobes are not well collimated. The easterly
outflow is thought to be a relatively young, small-scale outflow,
unresolved, from another source within the region. The polarimetry of
this region (Fig.~\ref{fig1}j) indicate that the magnetic field
aligns with the outflows -- the vectors are oriented north-south in
the north of the region, east-west in the east of the region and
northeast-southwest in the southwest of the region. The morphology of
the magnetic field is complex, with a plane of sky magnetic field
strength of $\sim$ 0.2 mG.

\subsection{NGC6334A}

NGC6334A is a giant H{\sc ii} region/molecular cloud complex and is
located close to the galactic plane at a distance of 1.7 kpc
\citep{neckel}. NGC6334-submm has two cores, one in the east, and a
brighter core (probably multiple, but not resolved within the JCMT
beam) in the west of the image. There is a CO bipolar outflow
associated with NGC6334A, which lies almost in the plane of the sky
\citep{sarma00} in a north-south direction \citep{depree}. There is a
ridge of material between the two cores in which the polarimetry
(fig.~\ref{fig1}k) suggests that the magnetic field is parallel to the
ridge. The magnetic field across the faint core has a
southeast-northwest direction, which curls round to east-west across
the ridge. The magnetic field across the brighter main core is more
complex, in the northeast of the core, the magnetic field has a
northeast-southwest direction, but towards the southwest of the core,
it is directed northwest-southeast, and in the very southwest of the
region there is a polarisation null. The abrupt change in the
orientation of the magnetic field may be due to two cores being
present, although unresolved in the JCMT beam. Previous observations
of OH maser emission have yielded estimates of the line of sight
magnetic field strength of $\sim$ 0.35 mG toward the source, with
maximum strengths of 0.5 mG \citep{mayo}. The polarimetry presented
here lead to estimates of the plane of the sky component of the field
strength to be $\sim$ 0.9 mG, consistent with the previous line of
sight measurements.

\subsection{RCrA}

The molecular complex Coronae Australis is $\sim$ 129 pc away from the
Sun \citep{marraco}, and is dominated by the centrally condensed core
centred near the emission line star R Cr A. \citet{nutter} have
recently studied this source in the submillimetre, and reveal three
submillimetre peaks within the main source, named SMM-1A,B and
C. SMM-1A is located in the south-eastern part on the core, with SMM-1B
and C being located in the north-east and northwest of the core.

R Cr A is one of the more evolved sources in the polarimetry sample,
and is classified as a Herbig Ae star \citep{marraco}. There is a CO
bipolar outflow with a position angle of $\sim$90\degr\, in the region
of R CrA \citep{walker85} but more recent molecular line mapping by
\citet{anderson} has cast doubt on whether R CrA is the driving
source. The polarimetry of the R CrA region (fig.~\ref{fig1}l)
indicate that the magnetic field is roughly parallel to the direction
of the outflow \citep[see also][]{clark}. There are also polarisation
nulls across this source, in two regions --- one in the north of the
source, and one in the south. A plane of sky field component of 1 mG
has been calculated.

\subsection{S140}

S140 is located in the Cepheus ring, at a distance of 900 pc
\citep{preibisch}. S140 has two outflows, one with a position angle of
$\sim$160\degr, which is bipolar in nature, and another smaller scale
outflow which has a position angle of $\sim$20\degr\,
\citep{preibisch}. The higher resolution K'-band data of
\citet{weigelt} show arc-like structures protruding from the northeast
of the source which are proposed to trace outflow cavities carved out
by material flowing away from S140 IRS1. The resolution of their data
is 240 milli-arcsec and covers an area of 13\arcsec $\times$
21\arcsec\, -- approximately one beam width of the SCUBA data
presented here. The SCUBA data (fig.~\ref{fig1}m) show that the
magnetic field vectors are ordered, in a north-south direction to the
western side of the source, whereas in the east, the magnetic field is
east-west orientated. OH Zeeman observations \citep{baudry97} have
revealed line of sight field estimates of +2.8 mG. CF estimates of the
plane of sky component of the field lead to strengths of $\sim$ 0.4
mG, which either indicates the magnetic field is mainly in the line of
sight, or, that the field is stronger on smaller scales. There is a
faint ridge of gas and dust extending from the east of the source,
curling northwards, which the magnetic field vectors seem to follow,
running parallel to it (in the plane of the sky). It may be possible
that the arc-like structure in the SCUBA image is related in some way
to the smaller scale outflow cavities seen by \citet{weigelt}.

\subsection{S146}
 
S146 is located at a distance of 5.2 kpc \citep{wu05}. A bipolar
molecular outflow, in a north-south direction (north lobe
blueshifted), driven by a star of spectral type O6.5 or earlier
(required to account for the ionisation of the H{\sc ii} region). In
the submillimetre data (fig.~\ref{fig1}n) there are two cores in a
north-south configuration. There is a ridge of gas and dust seemingly
connecting the two cores. The polarimetry appears to be almost
randomly distributed, with several polarisation nulls on the northern
core, and to the north and south of the southern core, although there
does not seem to be a relation between the intensity and polarisation
percentage. The large scatter of the polarimetry vector position
angles may suggest that the magnetic field is weak across this
region. It may also be explained if the magnetic field was
predominantly in the line of sight (the outflow is mainly in the line
of sight -- see table~\ref{tab:pol}), which can cause random
polarisation patterns and/or low polarisation percentages. The plane
of sky component of the field strength is calculated to be 0.1 mG,
which is one of the weakest measured for this sample.

\subsection{S157}

S157 is a diffuse nebula located towards the Cassiopeia-Perseus arm at
a distance of $\sim$ 2.5 kpc, and is surrounded by H{\sc ii} regions
and young open clusters \citep{shirley03}.  The submillimetre data
(fig.~\ref{fig1}o) show the region is only slightly more extended
than a point source, with bright submillimetre emission extending
southwards from the main core. The polarimetry of the core show that
the magnetic field vectors have an east-west direction to the south of
the core, but in the north, the vectors are aligned roughly
northwest-southeast. There are two regions of null polarisation ---
one in the northwest of the source, the other in the southeast. The
plane of sky field strength component is estimated to be $\sim$ 0.2
mG.

\subsection{W49}

W49 is in the galactic plane at a distance of 11.4 kpc \citep{gwinn},
and is one of the most luminous H{\sc ii} regions in the Galaxy
\citet[$\sim$ 10$^{7}$, ][]{smith78}. There are two bright cores in
this region (fig.~\ref{fig1}p) -- one in the northwest, W49N, and one
in the southeast, W49SE, with a third fainter source along the ridge
in between the brighter two cores, W49E. This ridge differs from those
previously mentioned, as it is much more extended and less
concentrated (fainter). There is a CO bipolar outflow from W49N, with
the redshifted lobe to the north and blueshifted to the south. The
outflow is almost in the line of sight \citep{scoville}. The
polarimetry vectors are aligned such that the magnetic field lines run
from one core to the other, parallel to the ridge. Across the W49N
core, the magnetic field (plane of the sky component) is estimated to
be less than 0.1 mG. The outflow is in the line of sight, however, and
so this may suggest that there is a larger line of sight component to
the magnetic field.

\section{Discussion}

\subsection{Comparison of Outflow \& B-field directions}

\begin{table*}
\centering
\caption{Outflow and magnetic field alignment. The columns are (1) the region name, (2) the variance in polarisation percentage, (3) the average polarisation percentage measured across the region, (4) the mean position angle of the magnetic field vectors, (5) the outflow direction, (6) the difference between the mean position angle of the magnetic field vectors and the outflow, (7) the opening angle of the outflow, (8) the inclination (from the plane of the sky) of the outflow, (9) notes and reference.} 
\label{tab:pol}
\begin{tabular}{l|c|c|c|c|c|c|c|c}
\hline
(1) & (2) & (3) & (4) & (5) & (6) & (7) & (8) & (9)\\
Region & $\bar{P}$ & $\sigma^{2}_{P}$ & $\bar{P.A.}$ & Outflow P.A. & $|\delta \theta| 
$ & Opening Angle & Inclination & Notes/References\\
  & \% & &  \degr\, & \degr\, & \degr\,& & & \\
   \hline
Cepheus A   &     1.7  & 1.3 & 52 & 45 &  7   & 15-20\degr\,     &  pos (62\degr)      & \citet{patel} \\
            &          &     &    & 90 & 38   & 60\degr\,        &  pos  (62\degr)	  & \citet{patel} \\
DR21(OH)    &     1.3  & 1.0 & 98 & 90 &  8   &                  &  pos		  & \citet{lai04} \\
GGD27       &     1.2  & 0.5 &105 & 20 & 85   & 1\degr\,         &  $\sim$pos          & \citet{gomez03,marti99} \\
GL2136      &     2.3  & 1.3 &102 &135 & 33   & 60\degr\,        &  pos                & \citet{kastner94} \\
GL2591      &     1.0  & 0.6 & 96 & 45 & 51   & $<$90\degr\,     &  45\degr\,          & \citet{hutawarakorn05} \\
            &          &     &    & 95 &  1   & $<$90\degr\,     &  45\degr\,          &  more collimated than the 45\degr\, outflow \\
GL437       &     1.0  & 0.6 & 50 &  0 & 50   & low              &  pos                & \citet{meakin05} \\
IRAS 20126+4104 & 0.8  & 0.4 & 74 &117 & 43   & 21\degr\,        &  pos                & \citet{hofner07} \\
                &      &     &    &171 & 83   & 70\degr\,        &  pos                & \citet{hofner07} \\
IRAS 20188+3928 & 0.9  & 0.3 & 90 &  0 & 90   &	            &  los                & \\
                &      &     &    & 45 & 45   &	            &  los                & \\
L1287       &     1.4  & 0.02& 75 & 45 & 30   & 50\degr\,        &  pos (60\degr)      & \citet{Umemoto00} \\
MonR2       &     1.2  & 0.7 & 64 &  0 & 64   &	            &  los                & \citet{xu06} \\
            &          &     &    & 45 & 19   &	            &  pos                &  \\
            &          &     &    & 90 & 26   &	            &  los                & \\
NGC6334A    &     0.9  & 0.8 & 96 &  0 & 84   &	            &  pos (10\degr)      & \citet{sarma00} \\
RCrA        &     2.5  & 7.0 & 95 & 90 &  5   & 60\degr\,        &  pos                & \citet{anderson} \\
S140        &     1.5  & 1.0 & 70 & 20 & 50   &	            &  pos                & \citet{preibisch} \\
            &          &     &    &160 & 90   &	            &  los                & \citet{minchin95} \\
S146        &     1.3  & 1.1 & 86 &  0 & 86   &	            &  los                & \citet{wu05} \\
W49NW       &     2.2  & 3.2 &119 &  0 & 61   & 30\degr\,        &  los                & \\
  \hline
\end{tabular}
\end{table*}
\normalsize

Theoretically, magnetic fields play an important role in the launching
and collimation mechanisms of outflows, and so it would be of interest
to see if any relationship between the observed magnetic field
direction on these large scales, and the jet/outflow axis
exists. Previous studies of the alignment of T-Tauri stars with the
local magnetic field \citep{menard} reveal a possible connection
between the strength of the CTTS jets and their orientation with
respect to the magnetic field. Interestingly, they conclude that the
CTTS's with jets align to the magnetic field, but as a whole sample
(i.e. both CTTS's with and without jets), the population is randomly
orientated with respect to the magnetic field, which they
say suggests either the influence of the magnetic field is dominant at
large scales (whole cloud) but largely decreases on the much smaller
scale of individual objects, or, the orientation of the CTTS's has
changed since they first formed.

\begin{figure*}
\includegraphics[height=8.5cm]{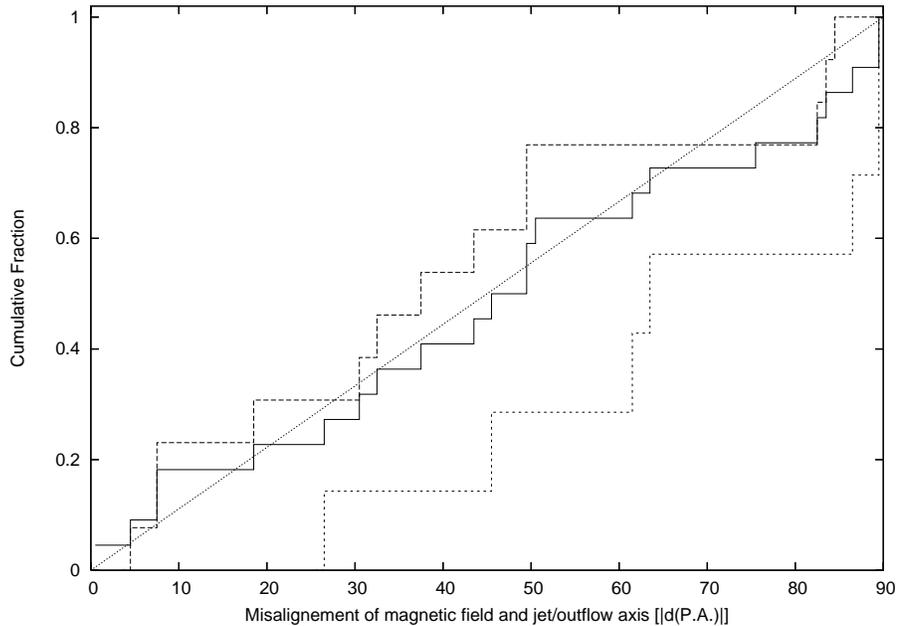}
\caption{Cumulative distribution function of the difference is position angles
between the weighted mean of the magnetic field vectors and the
jet/outflow axis. The long dashed histogram is for all sources with
outflow axes close to the plane of the sky ($i>45$\degr\,), the small
dashed histogram is for all sources with outflow axes close to the
line of sight ($i<45$\degr\,), the solid histogram is for the whole
sample. The dotted line is the function expected for an infinite
randomly oriented sample.}\label{misalign}
\end{figure*}

Table~\ref{tab:pol} shows the mean polarisation percentage, position
angles of the magnetic field and outflow, and the difference between
the magnetic field and outflow directions. The determined magnetic
field vectors are not true (i.e. undirectional) vectors. They have a
180\degr\, ambiguity, and as such have position angles of between
0\degr\, and 180\degr\,. The magnetic field direction for each region
is determined by calculating the weighted mean of the measured
vectors. The smallest difference between the magnetic field direction
and the outflow direction is assumed, and the results are plotted in
fig.~\ref{misalign}. Figure~\ref{misalign} is a cumulative distribution
function, which shows that, for the whole sample (excluding S157,
which as yet has no identified outflow, to the authors knowledge),
given the (weighted) mean position angle of the magnetic field
vectors, the magnetic fields within the sample appear randomly
oriented with respect to the the jet/outflow direction. The
Kolmogorov-Smirnov test reveals the whole sample has a 84.9\% chance of
being randomly orientated.

The sample has been split into two sub-samples -- those which have the
outflow axis mainly in the plane of sky (i.e. $i>45$\degr\, to the
line of sight), and those which have the outflow axis close to the
line of sight ($i<45$\degr\,). GL2591 is not included in either of
these sub-samples however, as the outflow axis lies close to $i \sim
45$\degr\,. Whilst all of the sources with very small differences are
in the plane of sky subsample, the cumulative distribution function
does not deviate significantly from a random distribution, and the
Kolmogorov-Smirnov test leads to a 57.2\% chance of random
orientation. However, the line of sight sub-sample, with no sources
showing good alignment between magnetic field and outflow axis, does
deviate significantly from the random distribution (a 36.6\% chance of
random orientation from the Kolmogorov-Smirnov test), suggesting a
relation between the misalignment of the (plane of the sky) field and
outflow when the outflow lies close to the line of sight. If an
alignment between the magnetic field and the outflow axis does exist
and the outflow is close to the line of sight, it would be difficult
to assign a direction to the magnetic field, and so misalignment
between the inferred field and outflow direction is more likely.

Previous mid-infrared spectropolarimetry \citep{aitken} reveal that
for their sample, overall, there is a large toroidal magnetic field
component within the molecular structures associated with embedded
young stellar objects. Their sample includes three sources from the
sample in this paper -- S140, MonR2 and GL2591. Their results for S140
and MonR2 agree with the findings of this paper -- namely that there
is a $\sim$ 40 \degr\, difference in alignment. However, they find the
magnetic field of GL2591 to be perpendicular to the 90\degr\, outflow,
whereas we find it in almost perfect alignment. This may be due to
either a changing magnetic field morphology on different scales, or
the magnetic field may have a stronger toroidal component closer in to
the protostar/disc. Compared to the submillimetre, the mid-infrared
emission comes from closer in towards the source, and it is higher
resolution (has a smaller beamsize).

The alignment analysis carried out here needs to be treated with
caution, for some regions, e. g. MonR2, S140, the mean position angle
may not accurately represent the field direction. MonR2, as with some
of the other sources, is an extended source, and the polarisation
pattern indicates abrupt changes in the direction of the magnetic
field. For sources such as these, the best alignment comparison may be
made using modal position angle values (in sufficient bins), or
possibly from the map itself.

Included in table~\ref{tab:pol}, where possible, are the opening
angles for the jets/outflows. Upon quick inspection, it is clear that,
with the information present, there appears to be no relation between
the degree of collimation and the alignment. GL2591 has a well
collimated jet (position angle $\sim$ 95\degr) which is almost
perfectly aligned to the magnetic field, whereas RCrA has an opening
angle of 60\degr\,, and again is well aligned to the magnetic
field. GGD27 has a well collimated jet with an opening angle of only
1\degr\,, but the axis is almost perpendicular to the magnetic
field. More information regarding the opening angles of such jets and
outflows is needed to enable a more detailed analysis.

\subsection{Field Morphology \& Submillimetre Emission}

A variety of continuum emission morphologies have been observed within
this sample. Whilst the majority (11 out of 16) of the sources have a
seemingly roughly spherical dense core (marginally resolved in the
JCMT beam), several of these sources also exhibit more extended
emission associated with the dense core (e. g. Cepheus A, DR21(OH),
IRAS 20188+3928). Of these sources, the polarisation percentage often
decreases with increasing intensity across the cores, whereas in the
extended emission the polarisation percentage often remains
stable. This may be due to the gas and dust being less dense in the
regions of the extended emission, or may be the magnetic field has a
simpler, less twisted morphology in these areas. Morphologically,
field lines both parallel (Cepheus A, W49) and perpendicular
(DR21(OH), S157) to the major axis of the extended emission are
observed.

This sample also included dense cores which deviate significantly from
spherical (at least in the plane of the sky, e.g. GL437, MonR2,
NGC6334A). Of these sources the magnetic field is observed to be
perpendicular to the major axis of the core in GL437 and RCrA. The
field morphology of MonR2 and NGC6334A are complex, with the fields
changing direction abruptly. This may be due to multiple cores
unresolved by the JCMT beam.

\subsection{Field Strength \& Cloud Support}

The plane of the sky magnetic field strengths have been calculated
based on signal-to-noise (3) clipped vectors. The fields range from
$<$0.1 mG (W49N) to 5.7 mG (Cepheus A), although the majority of them
are around 0.2--0.4 mG. Out of the three strongest magnetic fields
(Cepheus A, DR21(OH)N, RCrA), both DR21(OH)N and RCrA have field
morphologies that may indicate ambipolar diffusion (the field is
perpendicular to the major axis of the emission). Cepheus A, however,
has the strongest field, and the field is parallel to the major axis
of the extended emission. Previous polarimetric studies of W48 and
S152 \citep{hmpos}, reveal magnetic field strengths and polarisation
patterns consistent with the findings in this paper. W48 exhibits
depolarisation across the main core, whereas the polarisation
percentage remains the same across the candidate HMPO W48W. S152 is a
more complex region, consisting of ridges of gas and dust and the
S152SE core -- the candidate HMPO. The magnetic field is both parallel
and perpendicular to the ridges in places, but across the S152SE core,
the polarisation percentage remains at $\sim$ 8\% and the field is
perpendicular to the major axis of emission. The field strengths
across the candidate HMPOs were calculated to be 0.7 mG for W48W and
0.2 mG for S152SE, consistent with the magnetic field strengths
calculated for the star forming regions in this paper.

\section{Conclusions}

We present the largest (to date) sample of high mass star forming
regions observed using submillimetre polarimetry. We describe an
improved method of SCUBA polarimetry data reduction.  The magnetic
field geometries (from the intensity weighted vectors, projected onto
the plane of the sky) are presented, along with calculations of the
core masses, densities and field strengths.

The sources observed reveal a variety of morphologies for both the
continuum emission and the magnetic field. The majority of the sources
have dense cores that are unresolved in the beam, but some also have
extended emission associated with the cores. In several of the
regions, the polarimetry is uniform, suggesting ordered, relatively
strong magnetic fields. A decrease in polarisation percentage across
the cores is often seen, suggesting twisting or non-alignment of the
dust grains in these cores. The polarisation percentage often remains
stable across the extended emission, which may imply that in these
areas, the magnetic field has a simpler, less twisted morphology, or
it may be that as the extended emission is less dense, the dust grains
do not become misaligned to the magnetic field. We see field
morphologies perpendicular to the major axis of the continuum emission
(DR21(OH), RCrA, S157), which could suggest ambipolar diffusion for
these sources. It should also be noted that both of these regions
which have outflows, have magnetic fields that align well with the
outflow axes. We also see fields that are parallel to the extended
emission, e.g. Cepheus A, however, the field in this region also
aligns well with the outflow axis, and has the strongest magnetic
field strength in this sample. It is interesting in this case that the
extended emission is in the direction of the outflow.

An analysis of the mean position angle of the polarisation vectors and
the outflow axes has been carried out. Whilst the sample as a whole
has a cumulative distribution function similar to that expected if the
magnetic field and outflow axes were randomly orientated, if the
sample is broken down by inclination, into those predominantly in the
line of sight, and those predominantly in the plane of the sky, the
line of sight sample distinctly favours non-alignment, which would be
expected if the magnetic field had a large line of sight component
too, as it is then difficult to establish a magnetic field direction
from the polarimetry. The plane of the sky sample only shows a
marginal increase in alignment though. We discuss the caution needed
in interpretation of this analysis as the mean position angle of the
vectors may not be the best representation of the field direction for
all sources. In the sources where the field direction changes
abruptly, modal averages (in reasonable bin sizes) may be more
representative. Also, the polarimetry maps remain a very good way of
analysing the alignment.

\section*{acknowledgements}

The James Clerk Maxwell Telescope is operated by The Joint Astronomy
Centre on behalf of the Science and Technology Facilities Council
(STFC) of the United Kingdom, the Netherlands Organisation for
Scientific Research, and the National Research Council of Canada. The
authors acknowledge the data analysis facilities provided by the
Starlink Project which is run by CCLRC on behalf of STFC. The Program
ID's of the programs under which the data were obtained are: M98AU49,
M99AU03, M00BU09. RC acknowledges funding from the Science Foundation
Ireland, under grant 04/BRG/P02741.

\label{lastpage}

\end{document}